\documentclass[10pt,aps,prl,twocolumn,notitlepage,floatfix,showpacs,shownokeys,preprintnumbers,nofootinbib,tightenlines,longbibliography, superscriptaddress]{revtex4-1}
\pdfoutput=1
\usepackage[colorlinks=true,citecolor=blue,linkcolor=blue,breaklinks=true]{hyperref}
\usepackage{amsmath,amssymb,mathtools,tabu,bm}
\usepackage[T1]{fontenc}
\usepackage{textcomp}
\usepackage{lmodern}
\usepackage{enumerate}
\usepackage{cleveref}  
\usepackage{graphicx}   
\usepackage{commath}
\usepackage{slashed} 
\usepackage{comment}            
\usepackage{url}
\usepackage{color}
\usepackage{multirow}
\usepackage{placeins}
\usepackage[dvipsnames]{xcolor}
\usepackage{orcidlink}

\DeclareFontFamily{U}{mathx}{\hyphenchar\font45}
\DeclareFontShape{U}{mathx}{m}{n}{<-> mathx10}{}
\DeclareSymbolFont{mathx}{U}{mathx}{m}{n}
\DeclareMathAccent{\widebar}{0}{mathx}{"73}

\allowdisplaybreaks

\bibpunct{[}{]}{,}{n}{}{,}

\usepackage{tikz,xcolor,hyperref}
\definecolor{lime}{HTML}{A6CE39}
\DeclareRobustCommand{\orcidicon}{\hspace{-1mm}
	\begin{tikzpicture}
	\draw[lime, fill=lime] (0,0) 
	circle [radius=0.16] 
	node[white] {{\fontfamily{qag}\selectfont \tiny \,ID}};
	\draw[white, fill=white] (-0.0525,0.095) 
	circle [radius=0.007];
	\end{tikzpicture}
	\hspace{-3mm}
}

\foreach \x in {A, ..., Z}{\expandafter\xdef\csname orcid\x\endcsname{\noexpand\href{https://orcid.org/\csname orcidauthor\x\endcsname}
			{\noexpand\orcidicon}}
}

\begin{document}

\title{Role of Matter Inhomogeneity on Fast Flavor Conversion of Supernova Neutrinos}

\author{Soumya~Bhattacharyya\orcidlink{0000-0003-4960-8706}}
\email{soumyabhattacharyya475@gmail.com}
\affiliation{Institute of Physics, Academia Sinica, Taipei 115201, Taiwan}

\author{Meng-Ru Wu\orcidlink{0000-0003-4960-8706}}
\email{mwu@as.edu.tw}
\affiliation{Institute of Physics, Academia Sinica, Taipei 115201, Taiwan}
\affiliation{Institute of Astronomy and Astrophysics, Academia Sinica, Taipei 106319, Taiwan}
\affiliation{Physics Division, National Center for Theoretical Sciences, Taipei 106319, Taiwan}

\author{Zewei Xiong\orcidlink{0000-0002-2385-6771}}
\email{z.xiong@gsi.de}
\affiliation{GSI Helmholtzzentrum {f\"ur} Schwerionenforschung, Planckstra{\ss}e 1, D-64291 Darmstadt, Germany}

\preprint{}
\date{\today}

\begin{abstract}
\textcolor{black}{We study how a spatially varying matter potential $\lambda$, arising from neutrino–electron forward scattering, affects the onset, evolution, and nonlinear outcome of fast neutrino flavor conversions (FFCs) triggered by the presence of zero crossings in the angular distribution of the neutrino electron lepton number (ELN). 
We find that increasing the spatial variation rate of $\lambda$ can strongly influence FFC dynamics and even stabilize systems that are otherwise 
unstable}. 
Using stability analysis based solely on initial conditions, we identify for the first time a critical variation rate above which no FFC occurs even if the flavor instability exists. 
\textcolor{black}{Below this critical rate, a substantial $\lambda$ variation delays the onset of FFCs and quickly generates small-scale, incoherent features in the nonlinear regime, which leads to a similar coarse-grained outcome that eliminates the ELN crossing as in the homogeneous case.}
\textcolor{black}{Our findings 
emphasize the need to 
consider 
matter inhomogeneity in improved supernova models 
accounting for FFCs,  
and we propose simple analytical ways to incorporate this effect.} 
\end{abstract}

\maketitle 

\emph{\textbf{Introduction.---}} 
\textcolor{black}{Core-collapse supernovae (CCSNe) emit vast numbers of neutrinos~\cite{mirizzi2016supernova,burrows2020corecollapse,Mezzacappa2020physical,fischer2023neutrinos,janka2025longterm}, whose self-interactions~\cite{pantaleone1992neutrino,sigl1993general} can trigger FFCs~\cite{sawyer2005speed,sawyer2009multiangle,sawyer2016neutrino}—occurring at rates $\sim10^5$ times faster than vacuum oscillations—due to zero crossings in the ELN angular distribution shaped by intersecting $\nu_e$ and $\bar{\nu}_e$ fluxes~\cite{izaguirre2017fast,morinaga2022fast,dasgupta2022collective}.
Local neutrino kinetic studies neglecting spatial variation in the matter potential show that FFCs cause flavor depolarization, erasing flavor differences in neutrino fluxes~\cite{bhattacharyya2021fast,bhattacharyya2020late,wu2021collective,richers2021particle,richers2021neutrino,richers2022code,bhattacharyya2022elaborating,zaizen2023simple,zaizen2023characterizing,xiong2023evaluating,nagakura2023bgk,nagakura2023basic,cornelius2024perturbing,delfanazari2024systematic,xiong2024fast,fiorillo2024fast,xiong2024robust,shalgar2024neutrino,richers2024asymptotic,george2024evolution,fiorillo2025collective,liu2025asymptotic}, with major implications for CCSN dynamics and nucleosynthesis~\cite{xiong2020potential,fujimoto2023explosive,ehring2023fast,ehring2023fast2,nagakura2023roles,xiong2024fast,nagakura2024neutron,mori2025threedimensional,wang2025effect}; see~\cite{johns2025neutrino} for a recent review.}
Analyses of modern three-dimensional CCSN simulations revealed that FFC is more prominent in post-shock regions inside and above the proto-neutron star (PNS) within $O(0.1-1) \, s$~\cite{abbar2019occurrence,azari2019linear, glas2020fast,morinaga2020fast, capozzi2020fast,abbar2021characteristics, nagakura2021occurrence,akaho2023collisional}. 
During this space-time domain, the overall magnitude and spatial variation rate of the matter potential are significant, often matching or exceeding those of the neutrino self-interaction potential~\cite{chakraborty2011no,chakraborty2016self,xiong2023evolution,johns2025neutrino}, thus necessitating further study of FFC that incorporates the spatially inhomogeneous matter potential in neutrino quantum kinetics~\cite{sigl2022simulations}. 

In this \emph{Letter}, \textcolor{black}{we show how spatial inhomogeneity of the matter potential can strongly influence the onset, evolution and nonlinear outcome of FFCs}, depending on the strength of a dimensionless parameter $\gamma$, which roughly represents the ratio of the square of the neutrino self-interaction strength to the matter potential's spatial variation rate. 
\textcolor{black}{For $\gamma < 1$, strong matter variations can fully stabilize systems otherwise unstable to FFCs.} \textcolor{black}{Using local stability analysis~\cite{banerjee2011linearized,izaguirre2017fast} we derive an analytical condition for this stabilization based solely on the initial neutrino angular distribution. 
For substantial matter variation corresponding to $\gamma\gtrsim \mathcal{O}(10)$, while it delays the FFC onset and results in faster development of small-scale features, the coarse-grained outcome eliminates the ELN crossing as in cases where the matter potential is homogeneous.  
These results suggest that the matter variation effect can be conveniently included in subgrid models \cite{xiong2024robust,PhysRevD.109.083013} that use analytical prescriptions to incorporate FFCs into CCSN simulations. 
}

\emph{\textbf{Set-up \& Notation.---}}
To illustrate the phenomenon, we consider a dense, non-stationary, inhomogeneous, axially symmetric neutrino system within a small radial region $\Delta z$ along the radial axis (denoted as $z$ axis) inside a CCSN. 
Using the flavor-isospin convention~\cite{duan2006collective,chakraborty2016self2}, neutrinos and antineutrinos are treated together, with antineutrinos represented as particles with negative energy and negative phase-space densities. 
Approximating the \textcolor{black}{initial} neutrino angular distribution by a simplified two-beam model~\cite{chakraborty2016self2} with radial velocities $v_{+} = +1$ \textcolor{black}{for $\nu_e$} and $v_{-} = -1$ \textcolor{black}{for $\bar\nu_e$}, and neglecting vacuum oscillations, external forces, and momentum-changing collisions, the evolution 
\textcolor{black}{of these two modes} 
at any space-time point $(z, t)$ within $\Delta z$ is governed by the following equation of motion~\cite{sigl1993general,bhattacharyya2021fast}:
\begin{subequations}
\begin{align}
(\partial_t + \partial_{z} ) \mathbf{S}_{+} &= (\bm{\lambda}-2\mu \alpha \mathbf{S}_{-}) \times \mathbf{S}_{+}, \label{eq:eom_plus} \\
(\partial_t - \partial_{z} ) \mathbf{S}_{-} &= (\bm{\lambda}+2\mu \mathbf{S}_{+}) \times \mathbf{S}_{-}. \label{eq:eom_minus} 
\end{align}
\end{subequations}
In Eqs.~\eqref{eq:eom_plus}-\eqref{eq:eom_minus}, $\mathbf{S}_{\pm}[z, t]$ represents a three-dimensional flavor-space vector with basis $\{\hat{\mathsf{e}}_1, \hat{\mathsf{e}}_2, \hat{\mathsf{e}}_3\}$, describing the flavor composition of 

\textcolor{black}{the initial $\nu_e$ and $\bar\nu_e$} 
traveling 
with velocity $v_{\pm}$
at a specific space-time point $(z, t)$. 
Note that we will omit the explicit $z, t$ dependence of quantities unless needed for emphasis, shown as [...], e.g., $\mathbf{S}_{\pm}[z, t]$. 
The longitudinal component along $\hat{\mathsf{e}}_3$ is ${S}_{\pm}^{\parallel}$, while the transverse components in the $\hat{\mathsf{e}}_1-\hat{\mathsf{e}}_2$ plane are represented by a two-dimensional vector $\mathbf{S}^{\perp}_{\pm}$, with $ |\mathbf{S}_\pm^\perp|= \sqrt{1-(S^{\parallel}_{\pm})^2}$ that can be associated with the amount of flavor conversion. 
In Eqs.~\eqref{eq:eom_plus}-\eqref{eq:eom_minus}, $\bm{\lambda} = \lambda (0, 0, 1)$, where $\lambda(z) = \sqrt{2}G_F n_e(z)$ represents the matter potential due to neutrino 
forward scattering with inhomogeneous background electrons of density $n_e$, and  
$\mu = \sqrt{2} G_F n_{\nu_e}$ characterizes  the strength of the neutrino self-interaction potential originating from neutrino forward scattering among themselves. 
$\alpha$ is the initial neutrino number density ratio of the $v_-$ to $v_+$ mode, representing the depth of the zero crossing in the neutrino angular distribution for our two-beam model. 

Without loss of generality, we set $\alpha = 1/3$, $\mu = 1.5 \, \mathrm{cm}^{-1}$, and let $\lambda$ vary linearly in the $z$ direction

\textcolor{black}{according to $\lambda = m (z+90) \, \mathrm{cm}^{-1}$}. Here, $m$ represents the rate of spatial variation of $\lambda$ along the $z$-direction. 
We take different values of \textcolor{black}{$m = \{0, 10^{-6}, 10^{-4}, 10^{-2}, 6 \times 10^{-2}, 8 \times 10^{-2}, 9 \times 10^{-2}, 0.1, 0.3, 0.6\} \, \mathrm{cm}^{-2}$}. 
Using the finite volume version of the \texttt{COSE}$\nu$ solver~\cite{george2023cosenu}, Eqs.~\eqref{eq:eom_plus}-\eqref{eq:eom_minus} are solved up to \textcolor{black}{$t = t_f = 2.7 \, \mathrm{ns} $} with periodic boundary conditions over \textcolor{black}{$z \in \Delta z = (-90, +90) \, \mathrm{cm}$} discretized by \textcolor{black}{$18,000$} uniform grids, initializing \textcolor{black}{$S_{\pm}^{\parallel}[z, t = 0] = \sqrt{1-|\mathbf{S}^{\perp}_{\pm}[z,t=0]|^2}$, $\mathbf{S}_{\pm}^{\perp}[z, t=0] = 10^{-6} \, e^{-z^2/2w^2}$}, with \textcolor{black}{$w=2.24$~cm
\footnote{As the form of $\lambda$ introduces a discontinuity at the boundary, we only conduct our simulation up to $t_f$ to make sure that the flavor waves triggered by the central perturbation source of nonzero $\mathbf{S}_x$ does not cross the boundary. We have verified that a different choice of $\Delta z$ does not affect our result.}.
}

\begin{figure}[!t]
\includegraphics[width=25cm, height=7cm, keepaspectratio]{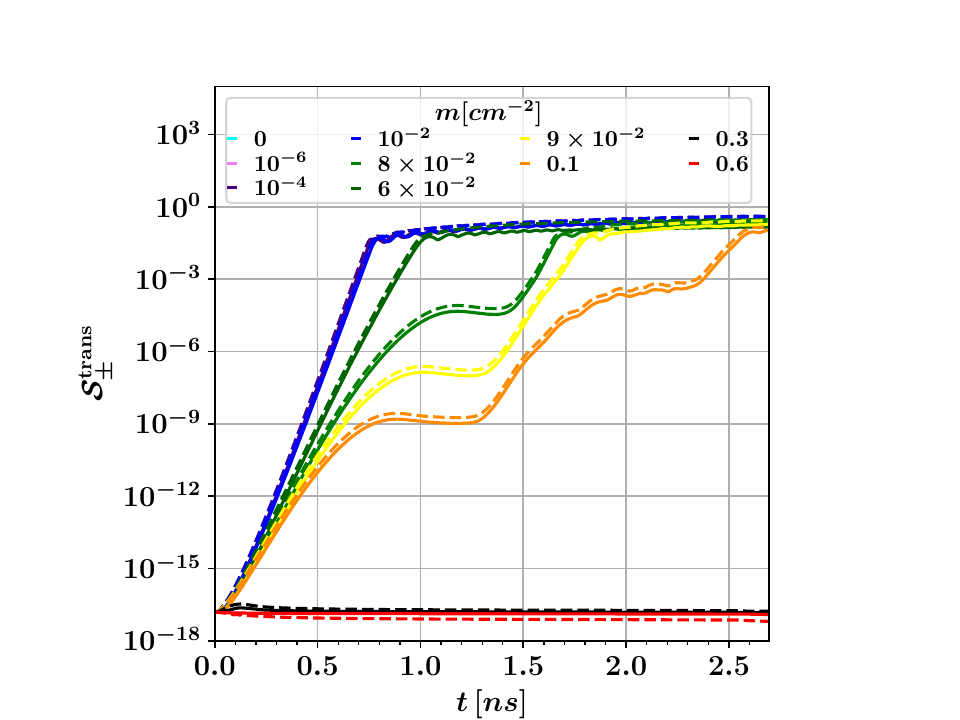}
\caption{
Time evolution of $S^{\mathrm{trans}}_{\pm}[t]$ for spatially varying $\lambda$ at different rates, denoted by $m$ (ranging from \textcolor{black}{$0$ to $0.6 \, \mathrm{cm}^{-2}$} as indicated in the legend) shown with different colors. 
Dashed and solid lines represent $S^{\mathrm{trans}}_{+}[t]$ and $S^{\mathrm{trans}}_{-}[t]$, respectively. 
For $m = 0$, the simulation includes a constant matter term $\lambda = 3 \times 10^{-3} \, \mathrm{cm}^{-1}$.
Lines with $m< 10^{-2}$~cm$^{-2}$ completely overlap with each other.
}
\label{fig:strans}
\end{figure}

\emph{\textbf{Effect of matter inhomogeneity.---}}
We define the measure of the overall FFCs in the simulation domain $\Delta z$ as $\mathcal{S}^{\mathrm{trans}}_{\pm}$, given by
\begin{equation}\label{eq:strans}
\mathcal{S}^{\mathrm{trans}}_{\pm}[t] =  \frac{1}{\Delta z} \int_{-\frac{\Delta z}{2}}^{\frac{\Delta z}{2}}  dz 
\left |\mathbf{S}_\pm^\perp[z,t]\right |.
\end{equation}
Its evolution in Fig.~\ref{fig:strans} reveals three regimes based on $m$:
\begin{itemize}
\item Small $m$ (\textcolor{black}{$0 < m~[\rm{cm}^{-2}] \lesssim 0.03$}): $\mathcal{S}_{\pm}^{\mathrm{trans}}$ behaves as in the absence of a matter term, leading to fast flavor depolarization. 
The flavor instability drives exponential growth of $\mathcal{S}^{\mathrm{trans}}_{\pm}$ for $t \lesssim O(1) \, \mathrm{ns}$ until it reaches an $O(1)$ value.  
\item Intermediate $m$ (\textcolor{black}{$0.03 \lesssim m~[\rm{cm}^{-2}] \lesssim 0.1$}): \textcolor{black}{
The growth of $\mathcal{S}^{\mathrm{trans}}_{\pm}$ is considerably slower, with the rate exhibiting multiple slopes or breaks before reaching an asymptotic value comparable to the $m = 0$ case. Larger $m$ further delays and suppresses the instability, so the asymptotic state is attained much later.}
\item Large $m$ (\textcolor{black}{$m~[\rm{cm}^{-2}] \gtrsim 0.1$}): \textcolor{black}{
$\mathcal{S}_{\pm}^{\mathrm{trans}}$ remains stable, indicating the absence of flavor conversion.}
\end{itemize}

\begin{figure*}[!t]
\includegraphics[width=0.37\textwidth]{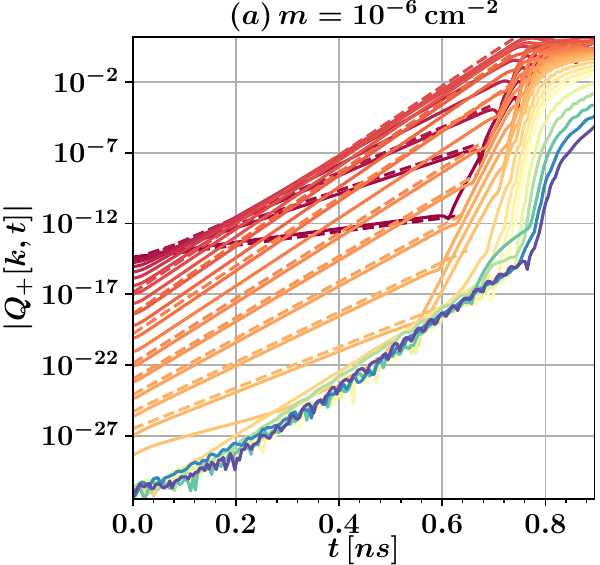}\hspace{1.0 cm}\includegraphics[width=0.55\textwidth]{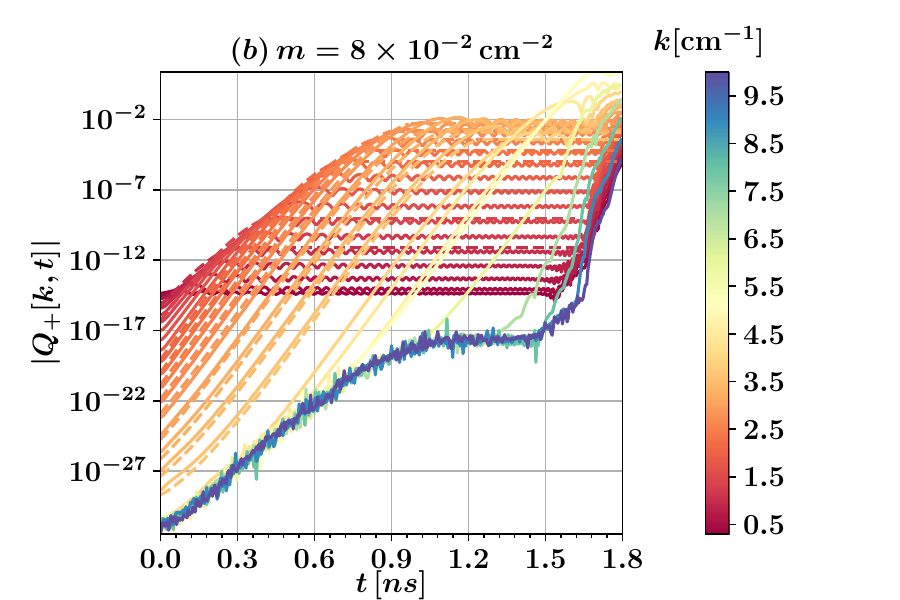}
\caption{\textcolor{black}{Numerical solutions for time evolution of $|Q_{+}[k, t]|$ across all $k$ modes (solid lines), color-coded according to the colorbar. Modes with $k = (0.3-3.6) \, \mathrm{cm}^{-1}$ are initially unstable.
Dashed lines indicate the analytical solutions (see text for details).The left and right panels correspond to {$m =  10^{-6} \, \mathrm{cm}^{-2}$ and $m =  8 \times 10^{-2} \,\mathrm{cm}^{-2} $}, respectively.}}  
\label{fig:qkt}
\end{figure*}

\emph{\textbf{Linear evolution.---}} \textcolor{black}{To study how matter inhomogeneity affects the early flavor evolution, we linearize Eqs.~\eqref{eq:eom_plus}–\eqref{eq:eom_minus}, assuming ${S}_{\pm}^{\parallel}[z, t] = 1$, and using the basis:}
\begin{equation}\label{eq:qdef}
Q_{\pm}[k, t] = \int dz \, e^{i(\lambda[z]t - kz)} \big(\mathbf{S}_{\pm}[z, t] \cdot (\hat{\mathsf{e}}_1 - i\hat{\mathsf{e}}_2)\big). 
\end{equation}
\textcolor{black}{This transformation yields time-dependent ODEs for each wavenumber $k$, removing the explicit $z$-dependence while encoding the spatial variation of $\lambda[z]$ through the slope $m$ in the form:}
\begin{equation}\label{eq:linear_ode}
\begin{split}
\partial_t \begin{pmatrix}
Q_{+} \\
Q_{-}
\end{pmatrix}
= \mathcal{A}_{2 \times 2}[t] \begin{pmatrix}
Q_{+} \\
Q_{-}
\end{pmatrix}, 
\end{split}
\end{equation}
where
\begin{equation}\label{eq:amatrix}
\mathcal{A}_{2 \times 2}[t] =
\begin{pmatrix}
-i(k-mt)+2i\mu\alpha & -2i\mu\alpha \\
2i\mu & i(k-mt)-2i\mu
\end{pmatrix}.
\end{equation}
\textcolor{black}{The magnitude of $Q_{\pm}[k, t]$ indicates the flavor conversion for a given $k$ mode at time $t$, with its approximate solution from Eq.~\eqref{eq:linear_ode} given by},
\begin{equation}\label{eq:qsol}
    \begin{split}
        Q_{\pm}[k, t]  \sim {Q}_{\pm}[k, 0] e^{-i \mu (1-\alpha)t-i\int_0^t \kappa[t] dt}
    \end{split},   
\end{equation}
where 
\begin{equation}\label{eq:kappa}
    \kappa[t] = \sqrt{(k-mt)^2-2(1+\alpha)\, \mu \, (k-mt)+\mu^2 (1-\alpha)^2}.
\end{equation}
\textcolor{black}{See Supplemental Material for the detailed derivation of Eq.~\eqref{eq:qsol} from Eq.~\eqref{eq:linear_ode}.}
\textcolor{black}{Eqs.~\eqref{eq:linear_ode}-\eqref{eq:kappa} show that for $m \neq 0$, $Q_{\pm}[k, t]$ evolves like the $m = 0$ case~\cite{chakraborty2016self2} but with $k$ replaced by $k_{\mathrm{eff}}[t] = k - mt$. 
Consequently, the linear growth rate governed by 
$\mathrm{Im}[\kappa[t]]$ becomes time-dependent, and the evolution of $Q_{\pm}[k, t]$ is now determined by the following time-integrated quantity,  
dubbed the ``growth integral'':  
\begin{equation}
    \Gamma_k[t] = \mathrm{Im}\left[\int_0^t \kappa[t] \, dt\right].
\end{equation}} In our two-beam model, $\mathrm{Im}[\kappa[0]]$ is nonzero within $k \in (k_{\mathrm{min}}, k_{\mathrm{max}})$, where $k_{\mathrm{min}} = \mu \, (1-\sqrt{\alpha})^2  \sim 0.27 \, \mathrm{cm}^{-1}$, $k_{\mathrm{max}} = \mu \, (1+\sqrt{\alpha})^2 \sim 3.73 \, \mathrm{cm}^{-1}$. It reaches a maximum of $\beta = 2 \mu \sqrt{\alpha} \sim 52 \, \mathrm{ns}^{-1}$ at $k = k_0 = \mu \, (1+\alpha) = 2 \, \mathrm{cm}^{-1}$ and symmetrically decreases to zero on both sides of $k_0$. 
\textcolor{black}{Eq.~\eqref{eq:kappa} shows that replacing $k$ by $k+mt$ leaves the shape of $\mathrm{Im}[\kappa[t]]$ unchanged, 
while the 
unstable $k$ range 
shifts toward higher $k$ modes over time.
The inverse timescale for a given $k$ mode with $k\gtrsim k_0$ that should experience unstable flavor growth can be estimated by the shift rate $r_s \sim m/(k_{\mathrm{max}}-k_0)=m/(2\mu\sqrt{\alpha}) \approx 17 (m/\mathrm{cm}^{-2}) \,\mathrm{ns}^{-1}$. 
Hence, one expects that when the shift rate is larger than the characteristic instability growth rate $\beta$, i.e., $\gamma\equiv\beta/r_s=4\mu^2\alpha/m\lesssim O(1)$, 
little flavor growth can develop for any given $k$ before being stabilized\footnote{\color{black}During the revision of this paper, two follow-up works that examined the impact of matter variation on collective flavor conversions appeared~\cite{Zaizen:2026wvj,Fiorillo:2026tee}, which also involve comparison of two relevant scales.
Subtle differences remain to be clarified, which is beyond the scope of this paper.}. 
This simple argument based on the comparison of two relevant scales, which depend solely on the initial ELN distribution,  suggests that FFCs should be completely suppressed when $m\gtrsim 3$~cm$^{-2}$,  consistent with Fig.~\ref{fig:strans}.
In fact, as will be shown later, $\gamma\sim\Gamma_k$ for $k$ modes that undergo flavor growths when matter inhomogeneity is substantial. 
Hence, for $\mathcal{S}_\pm^{\rm trans}$ to grow by $\mathcal{O}(10^8-10^{13})$ from its initial value, it requires $20\lesssim \gamma\lesssim 30$, i.e., {$m\sim \mathcal{O}(0.05)$}, also consistent with Fig.~\ref{fig:strans}. 
}
\begin{figure*}[!t]
\hspace{-0.75 cm}
\includegraphics[width=0.35\textwidth]{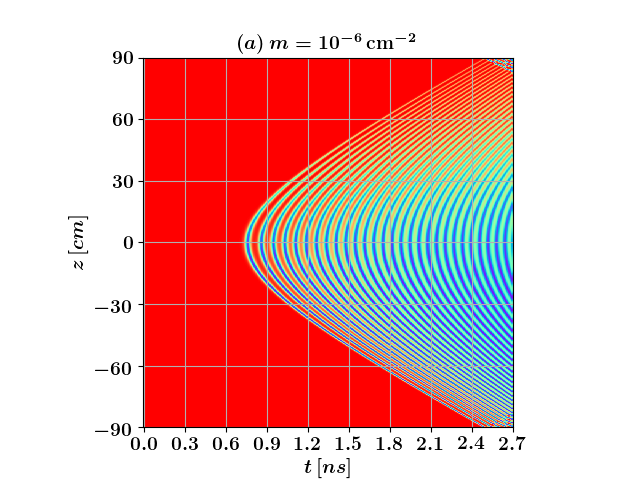}
\hspace{-0.75 cm}
\includegraphics[width=0.35\textwidth]{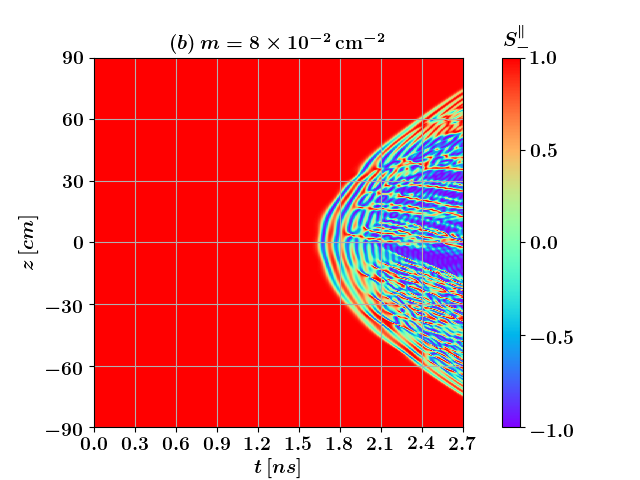}
\hspace{-0.75 cm}
\includegraphics[width=0.35\textwidth]{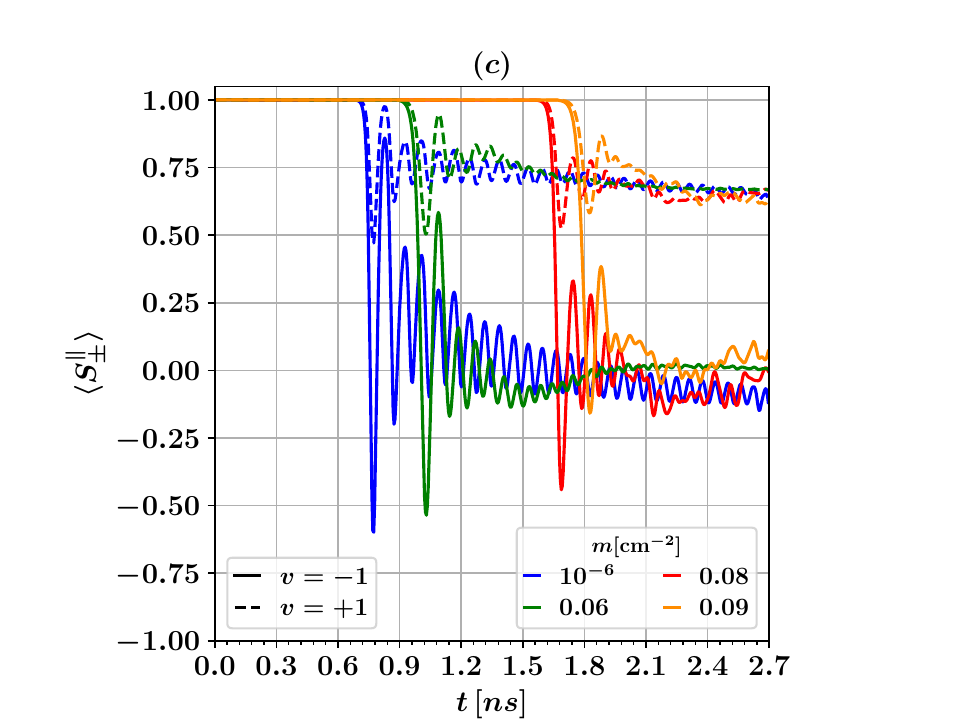}
\caption{Panels (a) and (b) show the evolution of {$S^{\parallel}_{-}$ in the $z-t$ plane}, corresponding to  $m = 10^{-6} \,\mathrm{cm}^{-2} $ and $m = 8 \times 10^{-2} \, \mathrm{cm}^{-2}$, respectively. \textcolor{black}{In panel (c), the time evolution of the spatially averaged $S^{\parallel}_\pm[z, t]$ for 4 different $m$ values are shown.
Solid and dashed lines represent the $+$ and $-$ modes, respectively.}
}
\label{fig:nonlinear}
\end{figure*}

\textcolor{black}{For small $m$ that corresponds to $\gamma \gg \mathcal{O}(30)$, 
$\mathrm{Im}[\kappa[t]]$ drifts in $k$-space at a negligibly small rate $r_s$ compared with the linear growth rates of the individual Fourier modes, which are of order $\beta$ or smaller. As a result, $\mathrm{Im}[\kappa[t]]$ remains nearly time-independent. 
Consequently, 
the evolution of all $k$-modes is governed by $\lim_{m \to 0} \Gamma_k[t] \sim \int_0^t \lim_{m \to 0} \mathrm{Im}[\kappa[t]] dt \sim \mathrm{Im}[\kappa[0]] t$, as if the matter potential were absent. 
Fig.~\ref{fig:qkt}(a) shows the temporal evolution of $Q_{+}[k, t]$ for $m=10^{-6}$~cm$^{-2}$, where only the initially unstable Fourier modes [$k \in (0.27-3.73) \, \mathrm{cm}^{-1}$] exhibit independent exponential growth up to $t \sim O(0.7) \, \mathrm{ns}$, in close agreement with the analytical prediction,  $|Q_{\pm}[k,t]| \sim |Q_{\pm}[k,0]|e^{\mathrm{Im}[\kappa[0]]t}$ (dashed lines).
Note that the growth for modes with $k>3.73 \,  \mathrm{cm}^{-1}$ at $t\lesssim 0.7$~ns arises from the higher order terms neglected in the linearized analysis of Eqs.~\eqref{eq:linear_ode}-\eqref{eq:kappa}.}

\textcolor{black}{For intermediate $m$ values ($\mathcal{O}(10) \lesssim \gamma \lesssim \mathcal{O}(30)$), 
we take $m=0.08$~cm$^{-2}$ ($\gamma=37.5$, $r_s \sim 0.03 \beta$) as an example and show the evolution of $|Q_+[k,t]|$ for different Fourier modes in Fig.~\ref{fig:qkt}(b). 
It shows modes with $k_{\rm min}<k<k_{\rm max}$ (red and darker orange curves), which are initially unstable, follow very similar growth trajectories with a time shift, and saturate to different values at different times. 
Those with $k>k_{\rm max}$  initially grow together due to the higher-order effect, but later follow the similar growth trajectories as the initially unstable modes before reaching the nonlinear regime. 
These behaviors can be understood by rewriting the growth integral as
\begin{equation}\label{eq:growth_int}
    \Gamma_k[t] = \frac{1}{m}\int_{x_{\rm min}}^{x_{\rm max}} \sqrt{R^2-x^2} dx,
\end{equation}
where $x = k-k_0-mt$ and $R = k_{\rm max}-k_0=2\mu\sqrt{\alpha}$.
The integration limits $x_{\rm max}$ and $x_{\rm min}$ correspond to the earliest and latest times when a $k$ mode becomes unstable (due to the shift of the unstable $k$ range), respectively. 
At large $t$ limit, i.e., $t\geq (k-k_{\rm min})/m$, 
\begin{equation}\label{eq:xlims}
x_{\rm min}=-R,~~ 
x_{\rm max}=
\begin{cases}
    k-k_0, &\textrm{for}~k_{\rm min}\leq k \leq k_{\rm max},\\
    R,&\textrm{for}~k > k_{\rm max}.
\end{cases}
\end{equation}
\textcolor{black}{Eqs.~\eqref{eq:growth_int} and \eqref{eq:xlims} show that, for any $k$, $x$ traces a semicircular trajectory of radius $R$ centered at $x = 0$ and the portion of the semicircle traversed by each $k$-mode over time represents its accumulated linear growth.} For the $k>k_{\rm max}$ modes, which are initially stable, they become unstable at later times and have the  same $\Gamma^0=\Gamma_k[t\rightarrow\infty]=\pi R^2/(2m)=2\pi\mu^2\alpha/m$ independent of $k$. 
For the initially unstable modes ($k_{\rm min}\leq k \leq k_{\rm max}$), larger $\Gamma_k[t\rightarrow\infty]$ is obtained for larger $k$. For $k=k_{\rm min}$, $k_0$, and $k_{\rm max}$, $\Gamma_k[t\rightarrow\infty]$ are 0, $\Gamma^0/2$, and $\Gamma^0$, resulting in  initial seed amplifications by 1, $\sim10^{13}$, and $\sim 10^{25}$, respectively. }

\textcolor{black}{
Another important aspect revealed by Fig.~\ref{fig:qkt}(b) is that all initially unstable modes fail to reach the nonlinear regime due to their initially small seeds. 
For the initially stable modes, they are able to reach the nonlinear regime because their initial growth is driven by nonlinear effects, which cannot be captured by linear analysis. 
\textcolor{black}{For instance, the growth of modes with $k \sim 4-5 \, \mathrm{cm}^{-1}$ (light orange and yellow curves) is first driven by nonlinear effects up to $t\sim 0.6 \, \mathrm{ns}$. 
Only at $t\gtrsim 0.6 \, \mathrm{ns}$, these modes become linearly unstable due to the shift of the unstable $k$ range, so that they continue to grow by $\sim 25$ orders of magnitude from $\mathcal{O}(10^{-22})$ to reach the nonlinear regime.}
This complexity indicates that for cases where matter inhomogeneity becomes substantial, effects beyond the linear analysis can in general be relevant.
}

\textcolor{black}{
Eqs.~\eqref{eq:growth_int} and \eqref{eq:xlims} also show that for large $m\gtrsim 2\pi\mu^2\alpha$, $\Gamma_k[t\rightarrow\infty]\lesssim\mathcal{O}(1)$ for all $k$, largely suppressing the flavor growths and ensuring long-term stability.}
\textcolor{black}{
Thus, a dense neutrino gas susceptible to fast instabilities becomes stabilized under a strongly varying matter potential, independent of the simulation space-time domain.
For completeness, we show the evolution of $|Q_+[k,t]|$ for the $m=0.6$~cm$^{-2}$ case in End Matter.
} 

\emph{\textbf{Nonlinear Evolution and Implications for CCSN Simulations.---}}
\textcolor{black}{
The above linear analysis shows that the flavor growth is unchanged (suppressed) with small (large) $m$ values. 
But for intermediate $m$ cases, 
while the flavor growths are delayed, they still reach the nonlinear regime. 
In Fig.~\ref{fig:nonlinear}(a) and (b), we show the spacetime evolution of $S_-^\parallel[z,t]$ for $m=10^{-6}$~cm$^{-2}$ and $m=0.08$~cm$^{-2}$ up to 2.7~ns, highlighting the nonlinear flavor wave structure. 
With $m=10^{-6}$~cm$^{-2}$, a coherent flavor wave pattern emerges as in cases without any matter inhomogeneity~\cite{martin2020dynamic}. 
When taking the intermediate $m=0.08$~cm$^{-2}$, substantial matter inhomogeneity not only delays the time reaching the nonlinear phase (see also Fig.~\ref{fig:strans}), but also leads to smaller and aperiodic spatial structures (see End Matter for the comparison of their Fourier spectra).
}
\textcolor{black}{
We speculate that
the reduced coherence 
arises from the shifting instability footprint, which causes the flavor evolution to be dominated by high-$k$ modes, thereby diffusing the neutrino flavor structure toward smaller spatial scales through enhanced mode-mode interactions. 
A complete understanding
is beyond the scope of this paper and is left for future work.}

\textcolor{black}{
In Fig.~\ref{fig:nonlinear}(c), we further show the time evolution of $\langle S_\pm^\parallel\rangle$, which are computed by averaging over the spatial domain encompassed within the envelopes shown in Fig.~\ref{fig:nonlinear}(a) and (b) for both $m=10^{-6}$ and $0.08$~cm$^{-2}$, as well as two additional cases with intermediate $m$ values of $0.06$ and $0.09$~cm$^{-2}$. 
The comparison shows that despite the differences in small scales and in the onset time, $\langle S_\pm^\parallel\rangle$ evolve toward similar quasistationary values for both velocity modes in all three intermediate $m$ cases. 
Notably, with the faster development of flavor depolarization, their $S_-$ modes actually approach coarse-grained flavor equipartition in a shorter amount of time. 
Moreover, the initial ELN crossing is removed after coarse-graining, as evidenced by $\langle S_-^\parallel \rangle \rightarrow 0$, just as in cases without matter inhomogeneity. 
}

\textcolor{black}{Thus, we propose the following simple method to include the effect of matter inhomogeneity in sub-grid models that aim to incorporate FFCs in global SN simulations. At each location where an ELN crossing is identified, in addition to estimate the instability growth rate  $\tau^{-1}$, one should also estimate the size of the unstable $k$ range, $\Delta k$,  as well as compute the matter potential gradient $m$. If $m\geq m_{\rm crit}\equiv\Delta k/\tau$, which corresponds to $\gamma\leq 1$, no flavor redistribution should be applied. On the other hand, when $m< m_{\rm crit}$, one may apply flavor redistribution as in the case without matter inhomogeneity. When applying our analysis of the inhomogeneous matter effect to simulation data of an artificially exploded, spherically-symmetric 18 $M_{\odot}$ SN model from Ref.~\cite{fischer2010protoneutron}, we estimate that the FFCs associated with shallow ELN crossings, whose characteristic depths are comparable to those observed in post-shocked regions in simulations that implemented the effects of FFCs in neutrino transport~\cite{xiong2024robust,wang2025effect}, should be suppressed (see End Matter for details). This suggests that the matter inhomogeneity effect discussed here can potentially influence the development and evolution of FFCs during the phase pertinent to shock revival. }

\emph{\textbf{Summary \& Outlook.---}} \textcolor{black}{We have presented a detailed study of how local matter inhomogeneity influences the development of FFCs. 
Assuming the spatial variation of the matter potential, characterized by $m = d\lambda/dr$, dominates over that of $\mu$, our numerical simulations demonstrate that matter inhomogeneity can delay or even completely suppress FFCs once $m$ exceeds a critical value. 
A local stability analysis rigorously shows that matter variation effectively introduces a time-dependent shift of the unstable region in Fourier space. 
The $e$-folding of the growth of an unstable Fourier mode is given by $\gamma\simeq {\rm Im}(\Omega)\Delta k/m$, where ${\rm Im}(\Omega)$ and $\Delta k$ (both of order $\mu$) are the characteristic instability growth rate and the width of the unstable branch in Fourier space for the dispersion relation solution in the absence of spatial inhomogeneity.} 

\textcolor{black}{
We find that for $\gamma \gg 1$ (small $m$), the evolution resembles the case without matter inhomogeneity, where rapid exponential growth occurs in the linear regime, followed by coherent wave-like oscillations in the nonlinear regime. 
For intermediate $\gamma \sim O(10)$ (intermediate $m$), FFCs are delayed or weakened, leading to slower growth in the linear regime. 
In the nonlinear regime, while flavor conversion experiences a delay and incoherent small-scale structures differently from the small $m$ case emerges, coarse-grained flavor conversion 
eliminates the initial ELN crossing 
as in the small $m$ case. 
For $\gamma\leq 1$ (large $m$), flavor conversion is fully suppressed.}

\textcolor{black}{Our results provide a quantitative criterion, via $m_{\mathrm{crit}}$ or $\gamma$, to assess the impact of matter inhomogeneity and 
suggest a way to include the matter variation effect into 
analytical prescriptions that are 
currently used in CCSN simulations taking into account FFCs, where shallow ELN crossings that may be affected by matter variations, particularly during the accretion phase critical for shock revival are present.} 

\textcolor{black}{Although our analysis used a simplified two–velocity-mode model, we have verified that the same conclusion extends to continuous angular distributions, which will be reported in a separate work.}
Moreover, we expect a similar or even stronger impact of matter inhomogeneity on slow and collisional instabilities due to their intrinsically weaker unstable growth rates~\cite{duan2010collective,johns2023collisional,johns2025neutrino}, warranting further exploration. 
In the same spirit, the potential impact of temporal variation of $\lambda$ may also be relevant. 
All these should be addressed to fully understand the role of neutrinos and their flavor oscillations in SN as well as in binary neutron star mergers, highlighting the profound implication of our work in this working direction.   

\begin{acknowledgements}
{\it Acknowledgments.---}We thank Sajad Abbar, Heng-Hao Chen, Huaiyu Duan, and Manu George for insightful discussions on the set-up and the results of this work. 
S. B. and M.R.W. acknowledge support of the National Science and Technology Council, Taiwan under Grant No.~111-2628-M-001-003-MY4, and the Academia Sinica (Project Nos.~AS-CDA-109-M11 and AS-IV-114-M04). 
M.R.W. also acknowledges support from the Physics Division of the National Center for Theoretical Sciences, Taiwan.
Z.X. acknowledges support of the European Research Council (ERC) under the European Union’s Horizon 2020 research and innovation program (ERC Advanced Grant KILONOVA No. 885281) and under the ERC Starting Grant (NeuTrAE, No. 101165138).
This work used ASGC (Academia Sinica Grid- computing Center) Distributed Cloud resources, which is supported by Academia Sinica.

The work is partially funded by the European Union. Views and opinions expressed are however those of the author(s) only and do not necessarily reflect those of the European Union or the European Research Council Executive Agency. Neither the European Union nor the granting authority can be held responsible for them.
\end{acknowledgements}

\bibliographystyle{apsrev4-1}
\bibliography{references_MRW}

\clearpage

\appendix
\onecolumngrid

\begin{center}
\textbf{\large End Matter}
\end{center}

\twocolumngrid

\begin{figure}[!h]
\includegraphics[width=0.45\textwidth]{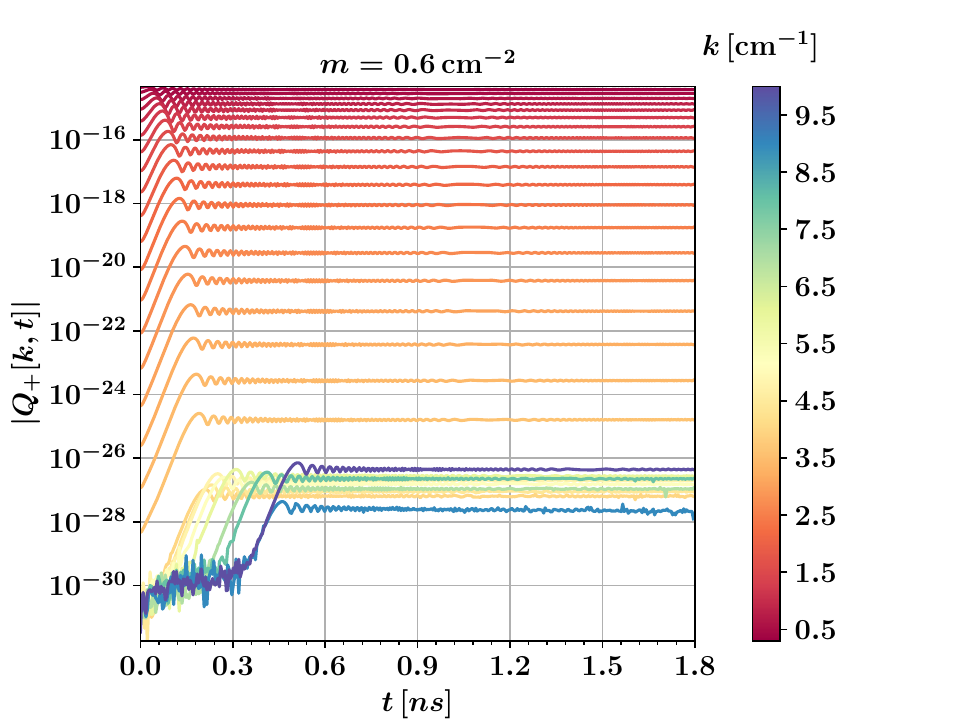}
\caption{{Time evolution of $|Q_{+}[k, t]|$ for all $k$ modes, with colors corresponding to the values indicated by the colorbar. Results are shown for the case $m = 0.6 \, \mathrm{cm}^{-2}$. }}
\label{fig:qkt_largem}
\end{figure}

\textcolor{black}{
\emph{\textbf{Flavor growth of Fourier modes with large $\bm{m}$.---}}For the large $m=0.6$~cm$^{-2}$ case, we show in Fig.~\ref{fig:qkt_largem} the flavor growths of different $k$ modes. 
Eq.~\eqref{eq:kappa} predicts $\Gamma_k[t\rightarrow\infty]\simeq7.85$ and the corresponding  flavor growth of $\simeq 2600$ for $k\gtrsim k_{\rm max}$, consistent with numerical results. 
For $k<k_{\rm max}$, the amount of flavor growth decreases due to the same reason discussed for the intermediate $m$ cases in the main text. 
Since the initial seeds are dominated by smaller $k$ modes, the limited flavor growths of the higher $k$ modes do not result in any apparent growth of $S_\pm^{\rm trans}$ shown in Fig.~\ref{fig:strans} of the main text.
}

\begin{figure}[!h]
\includegraphics[width=0.5\textwidth]{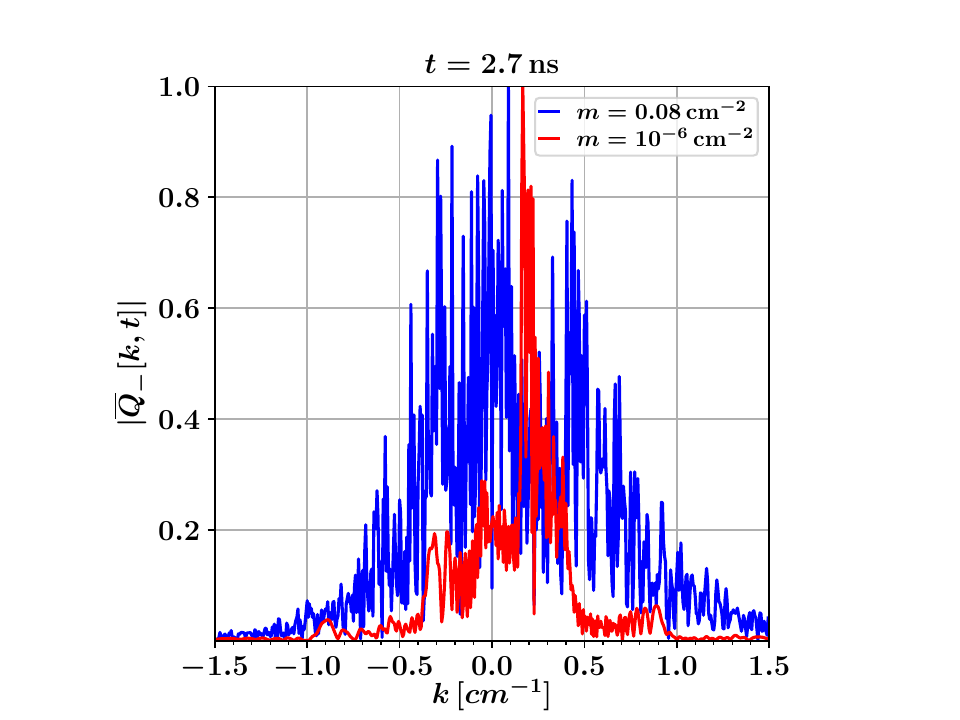}
\caption{{\textcolor{black}{$|\overline{Q}_{-}[k, t]|$ 
as a function of $k \in (-1.5, 1.5) \, \mathrm{cm}^{-1}$ at $t = 2.7 \, \mathrm{ns}$}. Different colors denote different values of $m$, as indicated in the legend: blue corresponds to $m = 0.08 \, \mathrm{cm}^{-2}$ and red to $m = 10^{-6} \,  \mathrm{cm}^{-2}$.}}
\label{fig:qk_nonlinear}
\end{figure}

\textcolor{black}{\emph{\textbf{Fourier spectra at the nonlinear regime.---}}}\textcolor{black}{Fig.~\ref{fig:nonlinear} of the main text shows that for the intermediate $m=0.08$~cm$^{-2}$ case, the smaller scale, incoherent features develop in the nonlinear regime, differently from that with a small $m=10^{-6}$~cm$^{-2}$. 
To quantify this, we additionally compute the direct Fourior transformation of $\mathbf{S}_\pm$ by  
\begin{equation}\label{eq:qdef_plain}
Q'_{\pm}[k, t] = \int dz \, e^{i kz} \big(\mathbf{S}_{\pm}[z, t] \cdot (\hat{\mathsf{e}}_1 - i\hat{\mathsf{e}}_2)\big), 
\end{equation}
and show in Fig.~\ref{fig:qk_nonlinear} the rescaled quantity 
\begin{equation}
\overline{Q'}_{-}[k, t] = \frac{Q'_{-}[k, t]}{{Q'}_{-}^{\mathrm{max}}[k, t]} , 
\end{equation}
where ${Q'}_{-}^{\mathrm{max}}[k, t]$ denotes the maximum value of $Q'_{-}[k, t]$ as functions of $k$ at $t=2.7$~ns for both $m$ values.  
The spectrum of the $m=10^{-6}$~cm$^{-2}$ case peaked at a particular value of $k\simeq 0.2$~cm$^{-1}$. 
In contrast, the spectrum becomes much wider and extends to higher $k$ modes with $m=0.08$~cm$^{-2}$, consistent with what discussed in the main text. 
} 

\begin{figure}[h]
\includegraphics[width=7cm, height=7cm, keepaspectratio]{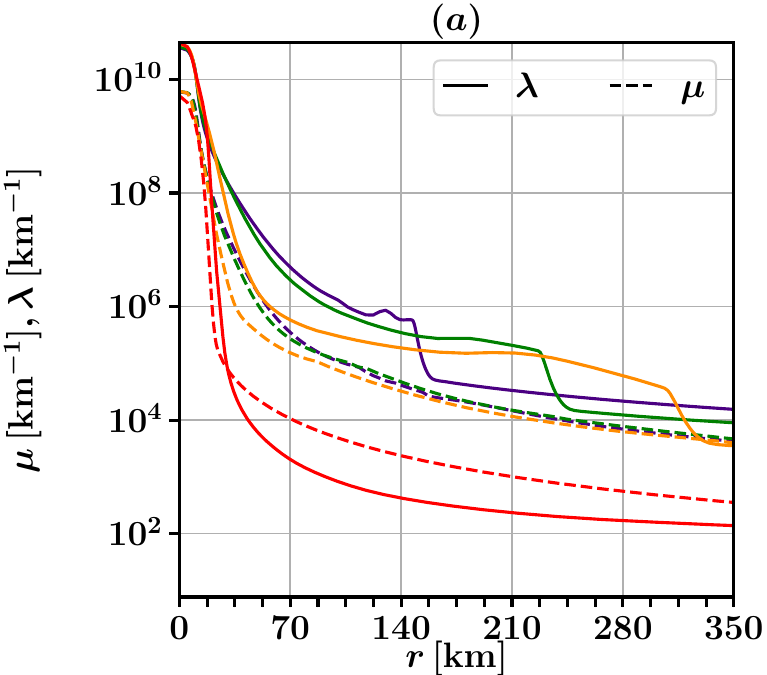}\includegraphics[width=7cm, height=7cm, keepaspectratio]{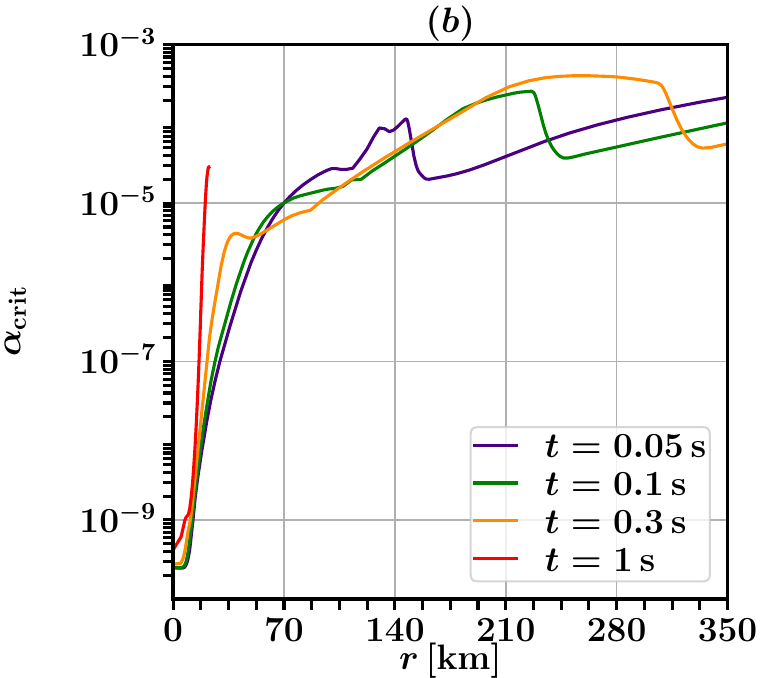}
\caption{Panel (a): Radial variation of $\mu$ (dashed) and $\lambda$ (solid) from an 18 $M_{\odot}$ SN model at four post-bounce phases: shock propagation ($t = 0.05 \, \mathrm{s}$), accretion ($t = 0.1 \, \mathrm{s},$ and $ 0.3 \, \mathrm{s}$) , and Kelvin-Helmholtz cooling ($t = 1 \, \mathrm{s}$), shown by different colors. 
Panel (b): The corresponding critical ELN zero crossing depth parameter $\alpha_{\mathrm{crit}} = m/(4\mu^2)$, derived from the local slope of $\lambda$. 
}
\label{fig:snprof}
\end{figure}

\emph{\textbf{Matter inhomogeneity in a realistic supernova.---}}We apply the analysis of inhomogeneous matter effect to simulation data of an artificially exploded spherically-symmetric 18 $M_{\odot}$ SN model from Ref.~\cite{fischer2010protoneutron} at four snapshots: $t = 0.05 \, \mathrm{s}$ (initial shock propagation), $t = 0.1 \, \mathrm{s}$ and $t = 0.3 \, \mathrm{s}$  (early and later accretion phases), and $t = 1 \, \mathrm{s}$ (PNS cooling phase). 
For each snapshot, we compute $m(r)=d\lambda(r)/dr$ and $d\mu(r)/dr$ based on $\lambda(r)$ and $\mu(r)$ shown in Fig.~\ref{fig:snprof}(a). 
We apply our analytical criteria to the radial range where $d\lambda(r)/dr \gtrsim d\mu(r)/dr$, which is valid in $r \leq 350$~km for $t = 0.05 \, \mathrm{s}$, $t = 0.1 \, \mathrm{s}$, and $t = 0.3 \, \mathrm{s}$, as well as $r \lesssim 30 \, \mathrm{km}$ for $t = 1 \, \mathrm{s}$. 

\textcolor{black}{
To estimate whether the realistic SN matter potential gradient $m(r)$ can be large enough to affect the development of FFCs in CCSNe, 
we assume that a possible ELN zero crossing at any location can be roughly represented by the simple two-beam model discussed in the main text, with  $\alpha$ denoting the ratio of the negative ELN on one side of the ELN crossing to the  positive ELN on the other side, which can be computed from realistic neutrino angular distributions. 
The corresponding fast instability growth rate 
$\tau_{\rm FFC}^{-1}$ and the size of the unstable $k$ range 
can then be estimated as $\tau^{-1}_{\rm FFC}\simeq\Delta k\simeq \beta=2\mu\sqrt{\alpha}$.
Thus, for a given $m(r)$, the condition $m\gtrsim m_{\rm crit}=\Delta k/\tau_{\rm FFC}\simeq 4\mu^2\alpha$ ($\gamma<1)$ for FFC suppression (see main text) allows to define the critical ELN crossing depth parameter 
$\alpha_{\mathrm{crit}}(r) = m(r)/(4\mu^2(r))$. 
At a given location $r$, the ELN crossing depth $\alpha(r)$ needs to be larger than $\alpha_{\mathrm{crit}}(r)$ for FFCs to develope.
}

Fig.~\ref{fig:snprof}(b) shows that $\alpha_{\rm crit}(r)$ generally increases with radius in all snapshots. 
For $t=0.05$~s, $t=0.1$~s and $t=0.3$~s,  $10^{-5}\lesssim\alpha_{\rm crit}\lesssim 10^{-3}$ in regions above the PNS.  
Since such $\alpha$ values are comparable to 

the shallow ELN zero crossing depths observed in post-shocked regions in simulations that implemented effect of FFCs in neutrino transport~\cite{xiong2024robust,wang2025effect}, this suggests that the matter inhomogeneity effect can influence the development and evolution of FFCs during the phase pertinent to shock revival. 
For $t=1$~s, $\alpha_{\rm crit} \lesssim 10^{-5}$ at regions where $d\lambda/dr \gtrsim d\mu/dr$, 
\textcolor{black}{indicating that the matter inhomogeneous effect is less relevant}. 

\onecolumngrid

\end{document}


\title{Supplemental Material to ``Role of Matter Inhomogeneity on Fast Flavor Conversion of Supernova Neutrinos''}

\author{Soumya~Bhattacharyya\orcidlink{0000-0003-4960-8706}}
\email{soumyabhattacharyya475@gmail.com}
\affiliation{Institute of Physics, Academia Sinica, Taipei 115201, Taiwan}

\author{Meng-Ru Wu\orcidlink{0000-0003-4960-8706}}
\email{mwu@as.edu.tw}
\affiliation{Institute of Physics, Academia Sinica, Taipei 115201, Taiwan}
\affiliation{Institute of Astronomy and Astrophysics, Academia Sinica, Taipei 106319, Taiwan}
\affiliation{Physics Division, National Center for Theoretical Sciences, Taipei 106319, Taiwan}

\author{Zewei Xiong\orcidlink{0000-0002-2385-6771}}
\email{z.xiong@gsi.de}
\affiliation{GSI Helmholtzzentrum {f\"ur} Schwerionenforschung, Planckstra{\ss}e 1, D-64291 Darmstadt, Germany}

\preprint{}
\date{\today}
\pacs{}

\maketitle

In this Supplemental Material, we provide a detailed derivation of the stability-analysis framework that incorporates the slowly varying matter potential used in the main text and examine the validity of two approximations behind Eq.~(6) therein. 

\section{Stability analysis including slowly varying matter potential}
We begin by linearizing Eqs.~(1a)-(1b) in the main text under the approximation $S^{\parallel}_{\pm}[z, t] = 1$, resulting in the following matrix equation expressed in terms of $S^{\perp}_{\pm}[z, t] = \mathbf{S}_{\pm}[z, t] \cdot (\hat{\mathsf{e}}_1 - i  \, \hat{\mathsf{e}}_2)$:
 \begin{sequation}\label{eq:linear_sys}
\begin{split}
i\begin{pmatrix}
\partial_t + \partial_{z}  & 0 \\
0 & \partial_t - \partial_{z} 
\end{pmatrix}
\begin{pmatrix}
S^{\perp}_{+} \\
S^{\perp}_{-}
\end{pmatrix}
= \begin{pmatrix}
\lambda[z] -2\mu\alpha & 2 \mu \alpha\\ 
-2 \mu & \lambda[z] + 2\mu
\end{pmatrix}
\begin{pmatrix}
S^{\perp}_{+} \\
S^{\perp}_{-}
\end{pmatrix}.
\end{split}
\end{sequation}   
We define a new basis, $Q_{\pm}[k, t]$, using the following relation:
\begin{sequation}\label{eq:qdef}
S^{\perp}_{\pm}[z, t] = \int_{-\infty}^{\infty} dk \, e^{i \, (k \, z-\lambda[z] \, t)} Q_{\pm}[k, t], 
\end{sequation}
where $\lambda[z] = m (z+90)$. It is important to note that Eq.~\eqref{eq:qdef} represents the Fourier transform of $S^{\perp}_{\pm}[z, t]$ in the comoving frame of $\lambda[z]$. Substituting Eq.~\eqref{eq:qdef} into Eq.~\eqref{eq:linear_sys}, we can write the time evolution of $Q_{\pm}[k, t]$ for each $k$ mode in terms of the following ordinary differential equation of matrix form:
\begin{sequation}\label{eq:qode}
\begin{split}
\partial_t \begin{pmatrix}
Q_{+} \\
Q_{-}
\end{pmatrix}
= \mathcal{A}_{2 \times 2}[t] \begin{pmatrix}
Q_{+} \\
Q_{-}
\end{pmatrix},
\end{split}
\end{sequation}
where

\begin{sequation}\label{eq:amatrix}
\mathcal{A}_{2 \times 2}[t] =
\begin{pmatrix}
-i(k-mt)+2i\mu\alpha & -2i\mu\alpha \\
2i\mu & i(k-mt)-2i\mu
\end{pmatrix}.
\end{sequation}

The $e^{-i \,\lambda[z] \, t}$ term in the basis transformation of Eq.~\eqref{eq:qdef} plays a key role in canceling the $z$-dependent $\lambda[z]$ term from the right-hand side of Eq.~\eqref{eq:linear_sys}. 
When the Liouville operator on the left-hand side acts on this term, it produces $(\partial_t \pm \partial_{z})e^{-i \,\lambda[z] \, t} = -i\big(\lambda[z]\pm \frac{d \lambda[z]}{dz} t \big) e^{-i \,\lambda[z] \, t}$. 
The first term cancels the $z$-dependent $\lambda[z]$ term on the right-hand side, leaving only the $\frac{d \lambda[z]}{dz} t$ term. 
For $\lambda[z]$ varying linearly in $z$ with slope $m$, this simplifies to $mt$, removing the explicit $z$-dependence in the equation of motion for $Q_{\pm}$. 
This step is crucial because it allows us to analyze the time evolution of $Q_{\pm}[k, t]$ for each Fourier mode $k$ independently using Eq.~\eqref{eq:qode}. 
To solve Eq.~\eqref{eq:qode}, we introduce a new basis $(\widetilde{Q}_{+}[k, t], \widetilde{Q}_{-}[k, t])$, which is connected to the original basis $({Q}_{+}[k, t], {Q}_{-}[k, t])$ through the following relation:
\begin{sequation}\label{eq:qtilde_transform}
\begin{pmatrix}
Q_{+}[k, t] \\
Q_{-}[k, t]
\end{pmatrix}
 = P_{2 \times 2}[t] \begin{pmatrix}
\widetilde{Q}_{+}[k, t] \\
\widetilde{Q}_{-}[k, t]
\end{pmatrix}
= (T_{X_1}[t], T_{X_2}[t])_{2 \times 2}\begin{pmatrix}
\widetilde{Q}_{+}[k, t] \\
\widetilde{Q}_{-}[k, t]
\end{pmatrix}.
\end{sequation}

In Eq.~\eqref{eq:qtilde_transform}, $T_{X_{1}}[t]$ and $T_{X_{2}}[t]$ are the $2 \times 1$-dimensional eigenvectors of the matrix $\mathcal{A}_{2 \times 2}[t]$ at time $t$ corresponding to the eigenvalues $X_{1}[t]$ and $X_{2}[t]$, respectively. 
From this point forward, we will omit the subscript $2 \times 2$ when referring to $2 \times 2$ matrices. 

From Eq.~\eqref{eq:amatrix}, the characteristic equation of the matrix $\mathcal{A}[t]$ can be solved to determine the explicit expressions for the eigenvalues ($X_{1, 2}[t]$) and eigenvectors ($T_{X_{1, 2}}[t]$) as functions of time $t$, as shown below: 
\begin{ssubequations}
    \begin{align}
     X_{1}[t] &= \mu (1-\alpha) + \kappa[t] \, \label{eq:x1},\\
     X_{2}[t] &= \mu (1-\alpha) - \kappa[t] \, \label{eq:x2},
     \end{align}
\end{ssubequations}
and
\begin{ssubequations}
    \begin{align}
     T_{X_{1}}[t] &=  \frac{1}{N_1[t]} \begin{pmatrix}
        \frac{-2\mu\alpha}{k_{\mathrm{eff}}[t]-2\mu\alpha-X_1[t]} \\
        1
     \end{pmatrix}, \label{eq:eigvec1}\\
     T_{X_{2}}[t] &=   \frac{1}{N_2[t]} \begin{pmatrix}
        \frac{-2\mu\alpha}{k_{\mathrm{eff}}[t]-2\mu\alpha-X_2[t]} \\
        1
     \end{pmatrix}. \label{eq:eigvec2}
     \end{align}
\end{ssubequations}
In Eqs.~\eqref{eq:x1}-\eqref{eq:eigvec2}, $\kappa[t]$ and $k_{\mathrm{eff}}[t]$ are defined as
\begin{ssubequations}
    \begin{align}
    \kappa[t] &= \sqrt{k^2_{\mathrm{eff}}[t]-2(1+\alpha)\, \mu \, k_{\mathrm{eff}}[t]+\mu^2 (1-\alpha)^2}, \label{eq:kappa} \\
    k_{\mathrm{eff}}[t] &= k-mt, \label{eq:keff}\\
    N_1[t] & =  \sqrt{\Bigg|\frac{2\mu\alpha}{k_{\mathrm{eff}}[t]-2\mu\alpha-X_1[t]}\Bigg|^2+1}, \label{eq:n1} \\
    N_2[t] & =  \sqrt{\Bigg|\frac{2\mu\alpha}{k_{\mathrm{eff}}[t]-2\mu\alpha-X_2[t]}\Bigg|^2+1}. \label{eq:n2}
    \end{align}
\end{ssubequations}

In Eq.~\eqref{eq:qode}, the dependence on $k$ in $\mathcal{A}[t]$ is expressed as $ k_{\mathrm{eff}}[t] = k-mt$. 
Therefore, in the solution for $Q_{\pm}[k, t]$ (or $\widetilde{Q}_{\pm}[k, t]$), $k$ appears as $ k_{\mathrm{eff}}[t] = k-mt$. 
From now on, we will simplify the notation by writing $Q_{\pm}[t]$ instead of $Q_{\pm}[k, t]$ (and $\widetilde{Q}_{\pm}[t]$ instead of $\widetilde{Q}_{\pm}[k, t]$), omitting the explicit dependence on $k$ and sometimes the time dependence when it is clear from context, particularly in differential equations. 
Moreover, the absolute value of $Q_{\pm}[t]$ will be represented as $|Q_{\pm}[t]|$. 
By applying the relation in Eq.~\eqref{eq:qtilde_transform} and assuming the existence of $P^{-1}[t]$, the equation of motion for $\widetilde{Q}_{\pm}[k, t] $ can be expressed as:
\begin{sequation}\label{eq:qtilde_ode}
\begin{split}
\partial_t \begin{pmatrix}
\widetilde{Q}_{+} \\
\widetilde{Q}_{-}
\end{pmatrix}
= \Bigg(-i\begin{pmatrix}
X_1[t] & 0\\
0 & X_2[t]
\end{pmatrix}-P^{-1}[t]\frac{dP[t]}{dt}\Bigg)
\begin{pmatrix}
\widetilde{Q}_{+} \\
\widetilde{Q}_{-}
\end{pmatrix}.
\end{split}
\end{sequation}
In Eq.~\eqref{eq:qtilde_ode}, the matrices $\frac{dP[t]}{dt}$ and $P^{-1}[t]$ are given by:
\begin{ssubequations}
\begin{align}
    \frac{dP[t]}{dt} &= 
\frac{1}{\kappa[t]}
\begin{pmatrix}
 \frac{2m\alpha\mu\big(\delta_{\mathrm{Im}[\kappa[t]], 0}+N_1[t]^2(1-\delta_{[\mathrm{Im}[\kappa[t]], 0]})\big) }{(N_1[t])^3(k_{\mathrm{eff}}[t]-2\mu\alpha-X_1[t])} & \frac{-2m\alpha\mu \big(\delta_{\mathrm{Im}[\kappa[t]], 0}+N_2[t]^2(1-\delta_{[\mathrm{Im}[\kappa[t]], 0]})\big)}{(N_2[t])^3(k_{\mathrm{eff}}[t]-2\mu\alpha-X_2[t])} \\
\frac{4m\alpha^2\mu^2\delta_{[\mathrm{Im}[\kappa[t]], 0]} }{(N_1[t])^3(k_{\mathrm{eff}}[t]-2\mu\alpha-X_1[t])^2} & \frac{-4m\alpha^2\mu^2\delta_{[\mathrm{Im}[\kappa[t]], 0]}}{(N_2[t])^3(k_{\mathrm{eff}}[t]-2\mu\alpha-X_2[t])^2} 
\end{pmatrix}, \label{eq:pmatrix} \\
P^{-1}[t] &= 
\frac{1}{\mathrm{det}[P[t]]} 
\begin{pmatrix}
\frac{1}{N_2[t]} & \frac{2\mu\alpha}{N_2[t](k_{\mathrm{eff}}[t]-2\mu\alpha-X_2[t])} \\
-\frac{1}{N_1[t]} & \frac{-2\mu\alpha}{N_1[t](k_{\mathrm{eff}}[t]-2\mu\alpha-X_1[t])}
\end{pmatrix} , \label{eq:pinv} 
\end{align}
\end{ssubequations}
where 
$\mathrm{det}[P[t]]$ is the determinant of the matrix $P[t]$ and $\delta_{[\mathrm{Im}[\kappa[t]], 0]}$ is the Kronecker delta function, defined as follows:
\begin{ssubequations}
\begin{align}
\mathrm{det}[P[t]] & = -\frac{\kappa[t]}{\mu N_1[t] N_2[t]}, \label{eq:detp} \\
\delta_{[\mathrm{Im}[\kappa[t]], 0]} & =  \begin{cases}
            1, &         \text{if } \mathrm{Im}[\kappa[t]]=0,\\
            0, &         \text{if } \mathrm{Im}[\kappa[t]] \neq 0.
    \end{cases} \label{eq:detp_sign}
\end{align}
\end{ssubequations}
If 
the elements of $P^{-1}[t]\frac{dP[t]}{dt}$ are negligible compared to $X_{1, 2}[t]$, an approximate solution for $\widetilde{Q}_{\pm}$ can be derived from Eq.~\eqref{eq:qtilde_ode} as
\begin{ssubequations}
    \begin{align}
        \widetilde{Q}_{+}[t] &= \widetilde{Q}_{+}[0]e^{-i \mu (1-\alpha)t-i\int_0^t \kappa[t] dt}, \label{eq:qtilde_plus} \\
        \widetilde{Q}_{-}[t] &= \widetilde{Q}_{-}[0]e^{-i \mu (1-\alpha)t+i\int_0^t \kappa[t] dt}. \label{eq:qtilde_minus} 
    \end{align}
\end{ssubequations}

Using Eqs.~\eqref{eq:qtilde_plus}-\eqref{eq:qtilde_minus} together with the transformation in Eq.~\eqref{eq:qtilde_transform}, one can obtain the solution for ${Q}_{\pm}[t]$ as a function of time $t$ for any given $k$ mode. 
Eqs.~\eqref{eq:qtilde_plus}-\eqref{eq:qtilde_minus} indicate that since $\mu$ and $\alpha$ are real, 
$\widetilde{Q}_{+}[t]$ for a specific $k$ mode can exhibit exponential growth over time, leading to fast flavor instability, only if $\kappa[t]$ for that mode becomes complex. 
In this case, $\widetilde{Q}_{-}[t]$ corresponds to the exponentially decaying solution and therefore does not contribute to the growth of flavor instability, allowing us to approximate $\widetilde{Q}_{-}[t] \sim 0$. 
Conversely, $\widetilde{Q}_{+}[t]$ represents the exponentially growing solution. 
Furthermore, by applying the transformation in Eq.~\eqref{eq:qtilde_transform}, the solution for ${Q}_{\pm}[t]$ for $k$ modes that exhibit exponential growth and drive fast flavor instability can be simplified as follows:

\begin{ssubequations}
   \begin{align}
       Q_{+}[t] & \sim {Q}_{+}[0] \Bigg(\frac{N_1[0]}{N_1[t]}\Bigg)\Bigg(\frac{N_{\mathrm{eff}}[0]}{N_{\mathrm{eff}}[t]}\Bigg)e^{-i \mu (1-\alpha)t-i\int_0^t \kappa[t] dt}, \label{eq:qsol_plus}\\
       Q_{-}[t] & \sim {Q}_{-}[0]\Bigg(\frac{N_1[0] }{N_1[t]} \Bigg)e^{-i \mu (1-\alpha)t-i\int_0^t \kappa[t] dt}, \label{eq:qsol_minus}
   \end{align}
\end{ssubequations}
where 
$N_{\mathrm{eff}}[t]$ is defined as: 
\begin{sequation}\label{eq:neff}
   N_{\mathrm{eff}}[t] = k_{\mathrm{eff}}[t]-2\mu\alpha-X_1[t]. 
\end{sequation}

Eqs.~\eqref{eq:qsol_plus}-\eqref{eq:qsol_minus} show that for $m \neq 0$, the evolution of $Q_{\pm}[t]$ resembles the $m = 0$ case, except that the wavenumber $k$ is replaced by a time-dependent $k_{\mathrm{eff}}[t] = k - mt$. 
This leads to time-dependent amplitude modulation proportional to $\frac{N_1[0]}{N_1[t]}\big(\frac{N_{\mathrm{eff}}[0]}{N_{\mathrm{eff}}[t]}\big)$ for $Q_{+}[t]$ (or $\frac{N_1[0] }{N_1[t]} $ for $Q_{-}[t]$) and a time-dependent frequency or growth rate $\sim \kappa[t]$, unlike the constant values in the $m = 0$ case. 
However, as we will demonstrate later in this supplemental material, for nearly all $m$ values and for the majority of the relevant portion of the linear evolution, the amplitude of $Q_{\pm}[t]$ varies negligibly over time compared to its frequency or growth rate.  
As a result, the time evolution of $Q_{\pm}[t]$ for any $k$ mode is primarily governed by the ``growth integral'' defined as:
\begin{sequation}
    \Gamma_k[t] = \mathrm{Im}[\int_0^t \kappa[t] \, dt].
\end{sequation}

In what follows, we will carefully examine how $|P^{-1}[t]\frac{dP}{dt}|$ compares with $|-i X_{1,2}[t]|$ as well as the variation of $N_1[t]$ and $N_{\rm eff}[t]$ for different regimes discussed in the main text. 

\subsection{Small $m$ regime}
For small $m$, with $mt \ll O(\mu)$, 
the leading-order terms in Eqs.~\eqref{eq:kappa}-\eqref{eq:n2} and Eq.~\eqref{eq:detp} are given by 
\begin{ssubequations}
    \begin{align}
    \kappa[t] & \sim \kappa[0]  \sim i\, O(\mu), \label{eq:approx_kappa} \\
    k_{\mathrm{eff}}[t] & \sim k  \sim O(\mu), \label{eq:approx_keff} \\
    N_1[t] & \sim N_1[0]  \sim O(1), \label{eq:approx_n1} \\
    N_2[t] & \sim N_2[0]  \sim O(1), \label{eq:approx_n2} \\
    \mathrm{det}[P[t]] & \sim \frac{-\kappa[0]}{\mu N_1[0] N_2[0]}.  \label{eq:approx_detp} 
    \end{align}
\end{ssubequations}

Consequently, the magnitudes of $|P^{-1}[t] \frac{dP}{dt}|$ and 
$|-i X_{1, 2}[t]|$ are
\begin{ssubequations}
    \begin{align}
    \bigg|P^{-1}[t] \frac{dP[t]}{dt}\bigg| & \sim \begin{pmatrix}
    O(\frac{m\alpha}{\mu}) & O(\frac{m\alpha}{\mu})\\
     O(\frac{m\alpha^2}{\mu}) & O(\frac{m\alpha^2}{\mu}) 
    \end{pmatrix}, \label{eq:approx_pdp}\\
    |-i X_{1, 2}[t]| & \sim O(\mu). \label{eq:approx_x12}
    \end{align}
\end{ssubequations}
With $\alpha < 1$ and small $m$, both $O(m\alpha/\mu) \ll O(\mu)$ and $O(m\alpha^2/\mu) \ll O(\mu)$, 
verifying that 
all elements of $|P^{-1}[t] \frac{dP[t]}{dt}|$ can be neglected compared to $|-iX_{1, 2}|$.

\subsection{Intermediate and large $m$ regimes}

For the intermediate- and large-$m$ regimes, Eq.~\eqref{eq:kappa} implies that for an initially unstable $k$ mode whose ${\rm Im}[\kappa[0]]>0$, ${\rm Im}[\kappa[t]]$ 
approaches zero as $mt$ increases due to the shift of the unstable $k$ range. 
At $t=t_c$, when $\kappa[t]=0$, i.e., 
the right-hand side of Eq.~\eqref{eq:kappa} vanishes, $\mathrm{det}[P[t]] = 0$, 
leading to the divergence of 
$P^{-1}[t]$. 
After that, $\kappa[t]$ for the same $k$ mode acquires a real, nonzero value that increases over time. 
For the initially stable $k>k_{\rm max}$ modes, this cancellation occurs twice: when the $k$ mode first becomes unstable and when it subsequently becomes stable again.

To verify the condition $|P^{-1}[t] \frac{dP[t]}{dt}| \ll |X_{1, 2}[t]|$ and assess whether the decoupled evolution equations, Eqs.~\eqref{eq:qsol_plus}-\eqref{eq:qsol_minus}, remain a good approximation for the system’s dynamics throughout the linear regime, we evaluate the ratio:
\begin{sequation}\label{eq:rij}
    R^{ij}[t] = \bigg|\frac{[P^{-1}[t] \frac{dP[t]}{dt}]^{ij}}{\kappa[t]}\bigg|
\end{sequation}
for 
\textcolor{black}{$k = 1.5, 3.0$, and $5 \, \mathrm{cm}^{-1}$ and show their evolution in 
Fig.~\ref{fig:rij} for the intermediate-$m$ regime, $m=0.08$~cm$^{-2}$, and the large-$m$ regime, $m=0.6$~cm$^{-2}$}. Here, $[P^{-1}[t] \frac{dP[t]}{dt}]^{ij}$ denotes the $(i, j)$-th element of the matrix $P^{-1}[t] \frac{dP[t]}{dt}$, with $i, j = 1, 2$. Note that we use $\kappa[t]$ instead of $X_{1,2}[t]$ in the denominator of $R^{ij}[t]$ because the constant term $\mu(1-\alpha)$ in $X_{1,2}[t]$ (Eqs.~\eqref{eq:x1}-\eqref{eq:x2}) does not affect the evolution of the amplitude of $Q_{\pm}[t]$ but only contributes to a phase. 
As shown in Fig.~\ref{fig:rij}, we find that $R^{ij}[t] \lesssim O(0.1)$ for all $i, j = 1, 2$ over most of the evolution. This justifies neglecting the term $P^{-1}[t] \frac{dP[t]}{dt}$ and using Eqs.~\eqref{eq:qsol_plus}-\eqref{eq:qsol_minus} to describe the linear dynamics, with errors below $O(10 \%)$ for the majority of the linear regime. 
Although $R^{ij}[t]$ exhibits brief divergences around $t_c$, 
Eqs.~\eqref{eq:qsol_plus}-\eqref{eq:qsol_minus} remain applicable, since the continuity of $Q_{\pm}[t]$ implied by Eq.~\eqref{eq:qtilde_ode} ensures that $Q_{\pm}[t]$ does not undergo significant discontinuous changes due to the transiently large values of $P^{-1}[t] \frac{dP[t]}{dt}$ that appear near $t \sim t_c$.
\begin{sfigure*}[!t]
	   \includegraphics[width=0.45\textwidth]{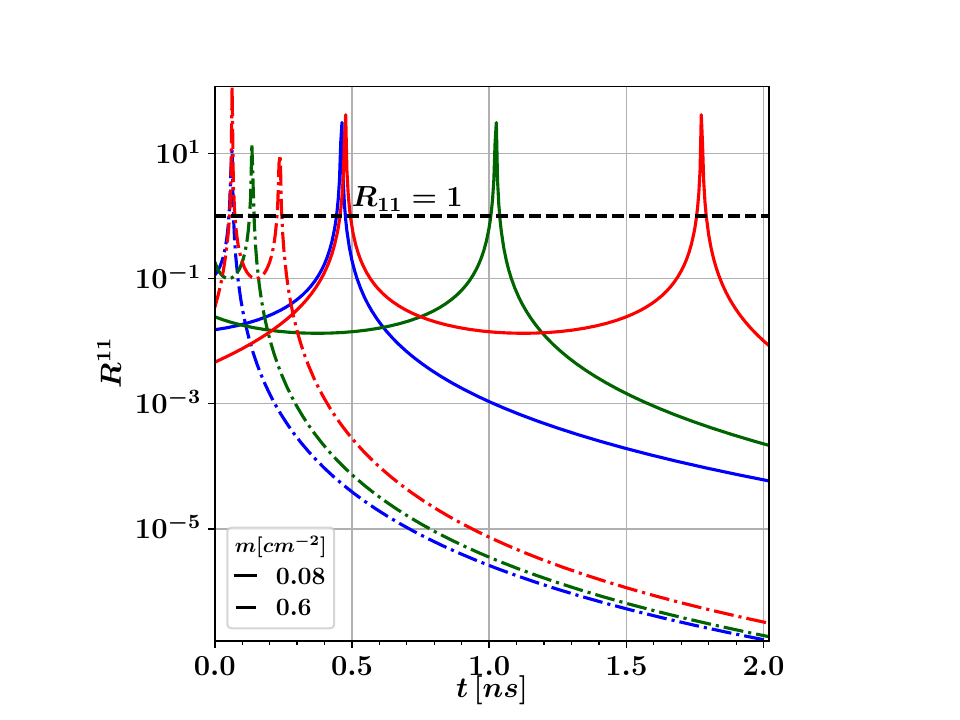}\hspace{1 cm}
        \includegraphics[width=0.45\textwidth]{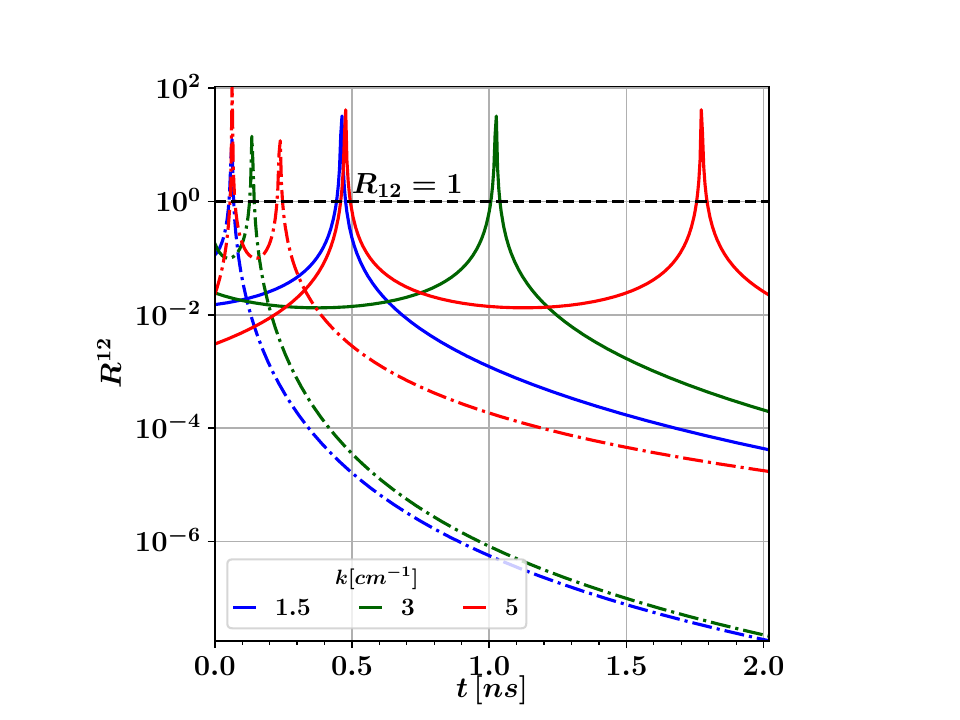}\\
        \vspace{0.6 cm}
        \includegraphics[width=0.45\textwidth]{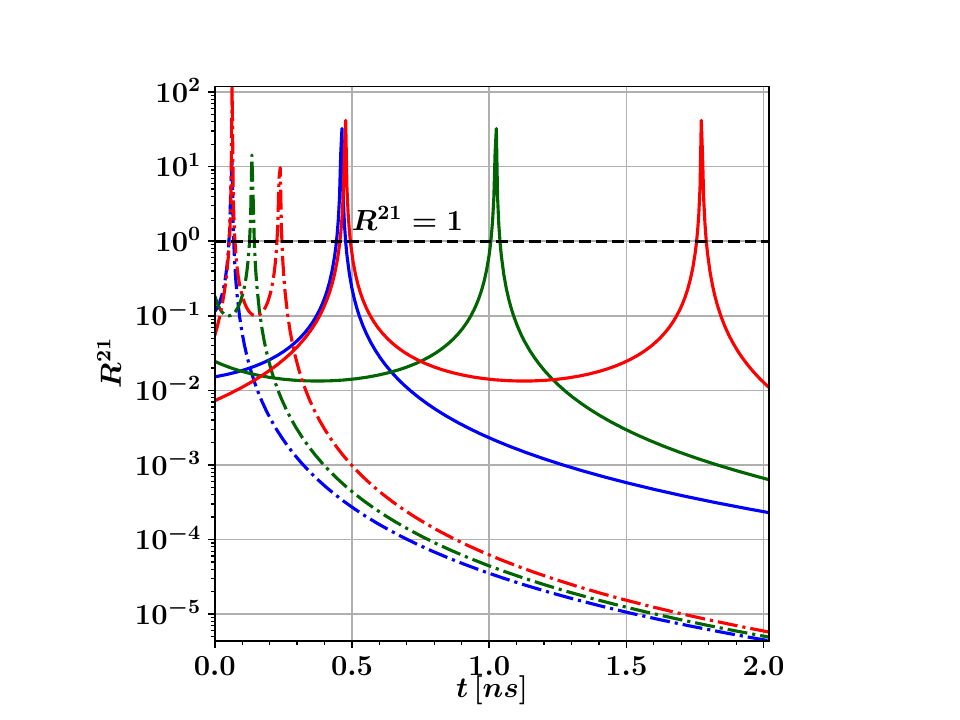}\hspace{1 cm}
        \includegraphics[width=0.45\textwidth]{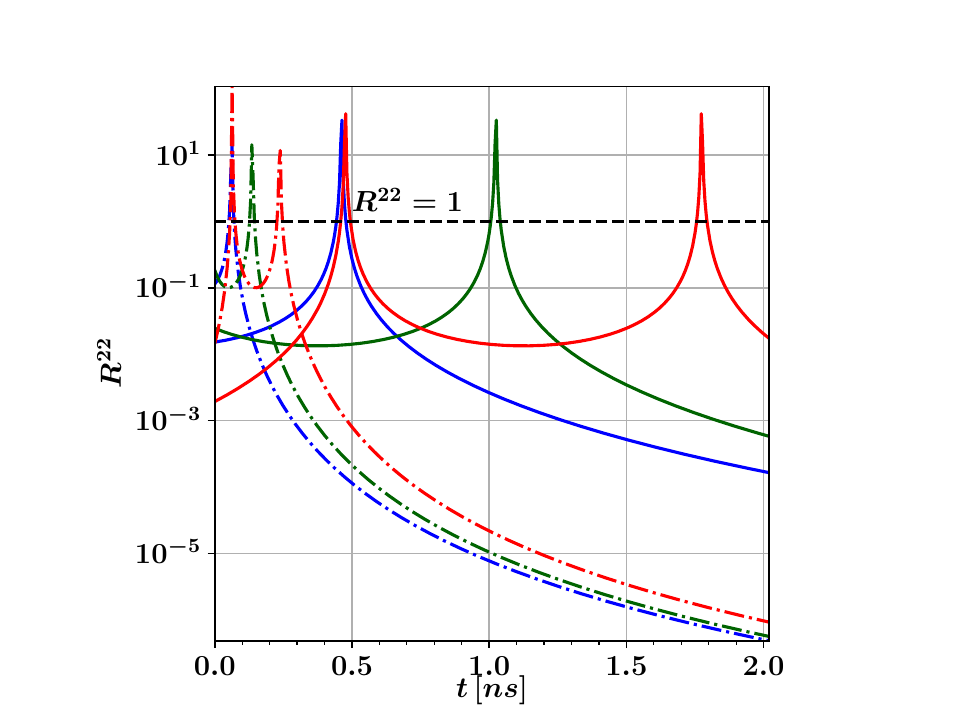}
\caption{{\sc :} Time evolution of $R^{11}$, $R^{12}$, $R^{21}$, and $R^{22}$, shown in the top-left, top-right, bottom-left, and bottom-right panels, respectively, up to $t = 2 \, \mathrm{ns}$ for \textcolor{black}{$k = 1.5, 3$, and $5 \, \mathrm{cm}^{-1}$, with $m = 0.08 \, \mathrm{cm}^{-2}$ (solid) and $m = 0.6 \, \mathrm{cm}^{-2}$ (dashed), respectively}. 
}
	\label{fig:rij}
\end{sfigure*}

As for the prefactors $N_1[0]/N_1[t]$ and $N_{\rm eff}[0]/N_{\rm eff}[t]$, their values in Fig.~\ref{fig:prefactors} vary by less than \textcolor{black}{$\mathcal{O}(10)$-$\mathcal{O}(100)$} for all $k$ modes, which do not significantly affect the exponential flavor growth behavior, justifying their omission in the discussion of the main text. \textcolor{black}{We have verified that this behavior persists in both the intermediate and large $m$ regimes.}

\begin{sfigure*}[!t]
	   \includegraphics[width=0.45\textwidth]{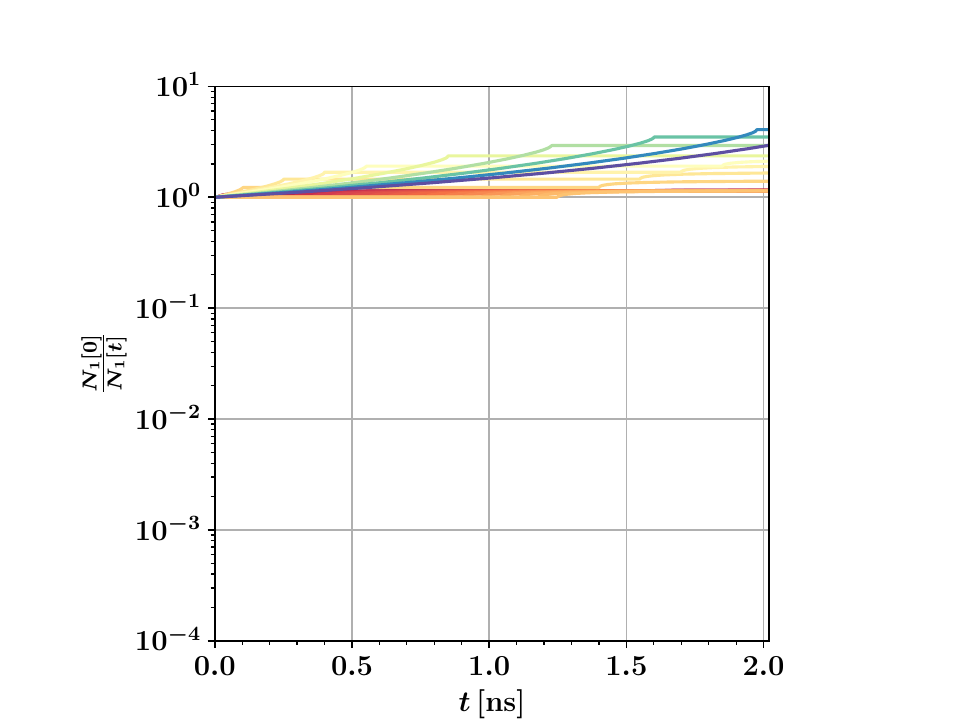}
       \hspace{1 cm}
        \includegraphics[width=0.45\textwidth]{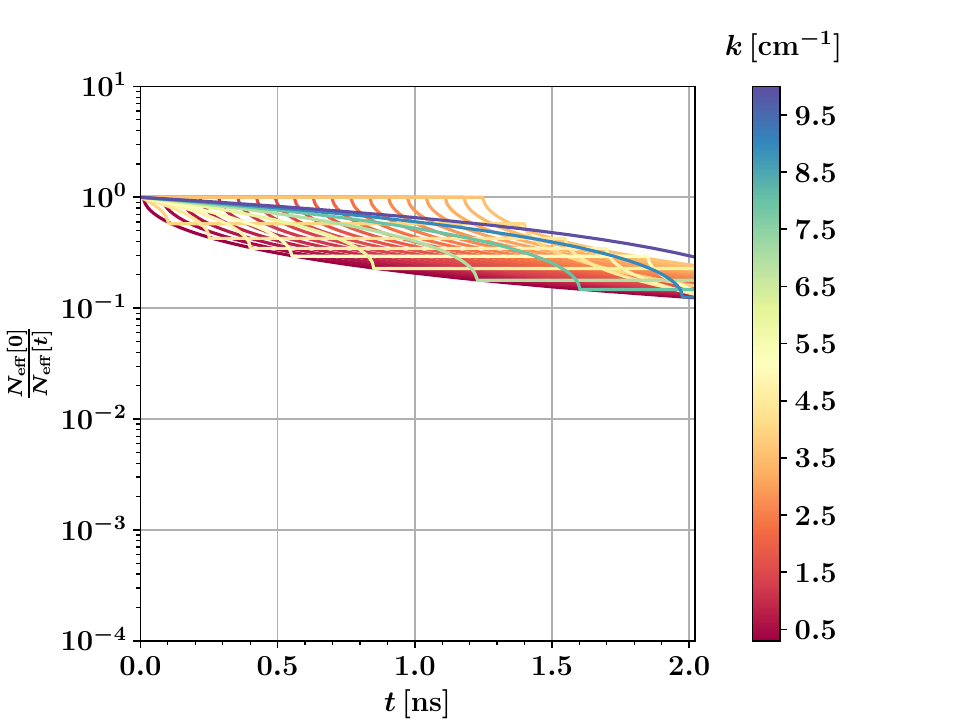}\\
        \vspace{0.6 cm}
        \includegraphics[width=0.45\textwidth]{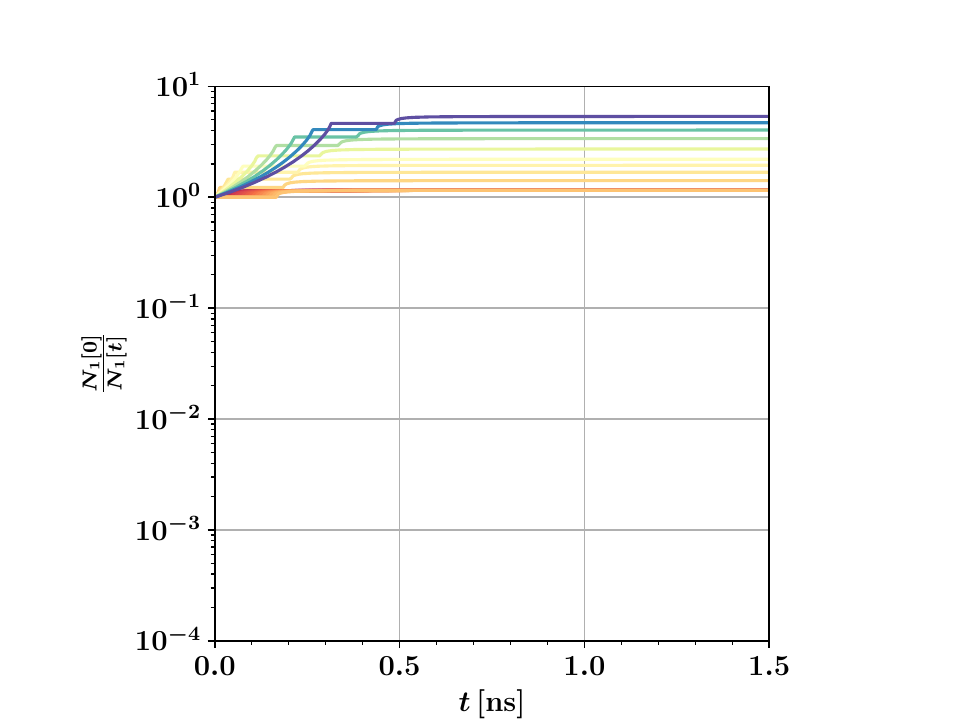}\hspace{1 cm}
        \includegraphics[width=0.45\textwidth]{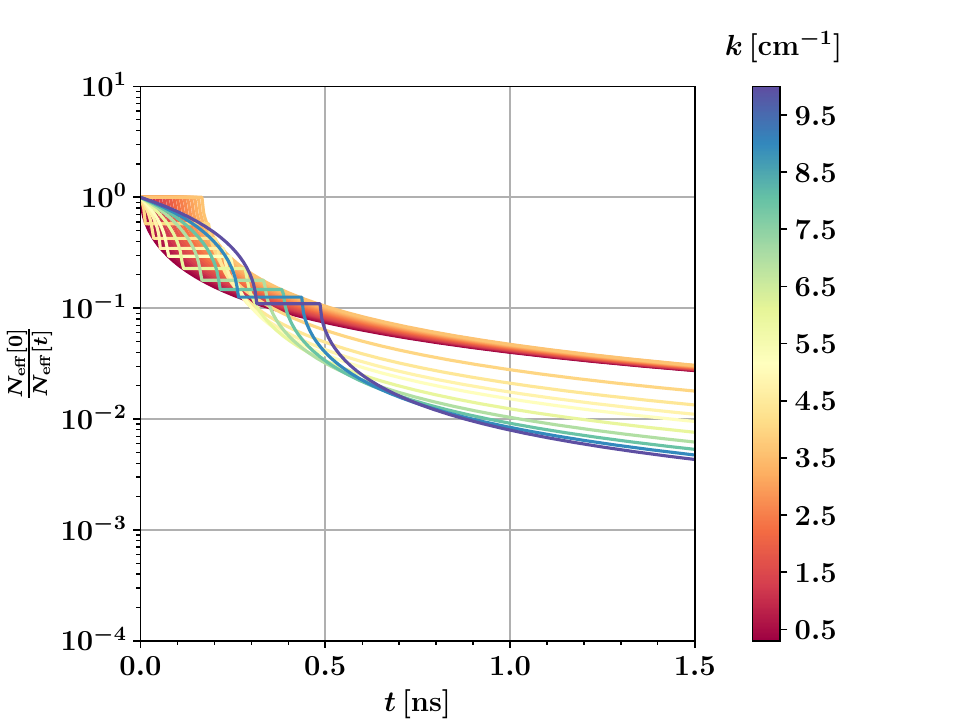}
\caption{{\sc :} Time evolution of $N_1[0]/N_1[t]$ (left panels) and $N_{\mathrm{eff}}[0]/N_{\mathrm{eff}}[t]$ (right panels) for all \textcolor{black}{$k$ modes}, with colors indicated by the colorbar \textcolor{black}{for the cases with $m = 0.08 \, \mathrm{cm}^{-2}$ (top row) and $m = 0.6 \, \mathrm{cm}^{-2}$ (bottom row).}}
\label{fig:prefactors}
\end{sfigure*}

\bibliographystyle{apsrev4-1}
\bibliography{references_MRW}